\documentclass[aps,prl,twocolumn,showpacs,preprintnumbers,amsmath,amssymb, superscriptaddress, bibnotes,longbibliography]{revtex4}
\usepackage[utf8]{inputenc}
\usepackage{graphicx}
\usepackage{dcolumn}
\usepackage{bm}
\usepackage[colorlinks,linkcolor=blue,anchorcolor=blue,citecolor=blue,urlcolor=blue,filecolor=blue,menucolor=blue,runcolor=blue]{hyperref}
\usepackage{ulem}

\makeatletter
\@ifundefined{textcolor}{}
{%
	\definecolor{BLACK}{gray}{0}
	\definecolor{WHITE}{gray}{1}
	\definecolor{RED}{rgb}{1,0,0}
	\definecolor{GREEN}{rgb}{0,1,0}
	\definecolor{BLUE}{rgb}{0,0,1}
	\definecolor{CYAN}{cmyk}{1,0,0,0}
	\definecolor{MAGENTA}{cmyk}{0,1,0,0}
	\definecolor{YELLOW}{cmyk}{0,0,1,0}
}

\usepackage{color}

\newcommand{\beq}{\begin{eqnarray}}
\newcommand{\eeq}{\end{eqnarray}}
\newcommand{\ys}[1]{\textcolor{black}{#1}}
\newcommand{\KVS}{{KV$_3$Sb$_5$}}
\newcommand{\AVS}{{$A$V$_3$Sb$_5$}}

\makeatother
\usepackage[english]{babel}

\begin{document}

\title{Soft Phonon Charge-Density Wave Formation in the Kagome Metal KV$_3$Sb$_5$}

\author{Yifan Wang}
\thanks{These authors contributed equally.}
\affiliation{Center for Correlated Matter and School of Physics, Zhejiang University, Hangzhou 310058, China}

\author{Chenchao Xu}
\thanks{These authors contributed equally.}
\affiliation{School of Physics, Hangzhou Normal University, 310036 Hangzhou, China}

\author{Zhimian Wu}
\affiliation{Hefei National Research Center for Physical Sciences at the Microscale, University of Science and Technology of China, Hefei, Anhui 230026, China}

\author{Huachen Rao}
\affiliation{Hefei National Research Center for Physical Sciences at the Microscale, University of Science and Technology of China, Hefei, Anhui 230026, China}



\author{Zhaoyang Shan}
\affiliation{Center for Correlated Matter and School of Physics, Zhejiang University, Hangzhou 310058, China}

\author{Yi Liu}
\affiliation{School of Physics, Zhejiang University, Hangzhou 310058, China}
\affiliation{Department of Applied Physics, Key Laboratory of Quantum Precision Measurement of Zhejiang Province, Zhejiang University of Technology, Hangzhou, China}

\author{Guanghan Cao}
\affiliation{School of Physics, Zhejiang University, Hangzhou 310058, China}
\affiliation{State Key Laboratory of Silicon and Advanced Semiconductor Materials,Zhejiang University, Hangzhou 310058, China}
\affiliation{Institute of Fundamental and Transdisciplinary Research, Zhejiang University, Hangzhou 310058, China}

\author{Michael Smidman}
\affiliation{Center for Correlated Matter and School of Physics, Zhejiang University, Hangzhou 310058, China}

\author{Ming Shi}
\affiliation{Center for Correlated Matter and School of Physics, Zhejiang University, Hangzhou 310058, China}

\author{Huiqiu Yuan}
\affiliation{Center for Correlated Matter and School of Physics, Zhejiang University, Hangzhou 310058, China}
\affiliation{Institute for Advanced Study in Physics,Zhejiang University,Hangzhou 310058,China}
\affiliation{State Key Laboratory of Silicon and Advanced Semiconductor Materials,Zhejiang University, Hangzhou 310058, China}
\affiliation{Institute of Fundamental and Transdisciplinary Research, Zhejiang University, Hangzhou 310058, China}

\author{Tao Wu}
\affiliation{Hefei National Research Center for Physical Sciences at the Microscale, University of Science and Technology of China, Hefei, Anhui 230026, China}

\author{Xianhui Chen}
\affiliation{Hefei National Research Center for Physical Sciences at the Microscale, University of Science and Technology of China, Hefei, Anhui 230026, China}

\author{Chao Cao}
\email{ccao@zju.edu.cn}
\affiliation{Center for Correlated Matter and School of Physics, Zhejiang University, Hangzhou 310058, China}

\author{Yu Song}
\email{yusong\_phys@zju.edu.cn}
\affiliation{Center for Correlated Matter and School of Physics, Zhejiang University, Hangzhou 310058, China}

\begin{abstract}
A range of of unusual emergent behaviors have been reported in the charge-density wave (CDW) state of the $A$V$_3$Sb$_5$ ($A=~$K, Rb, Cs) kagome metals, including a CDW formation process without soft phonons, which points to an unconventional CDW mechanism. Here, we use inelastic x-ray scattering to show that the CDW in KV$_3$Sb$_5$ forms via phonons that soften to zero energy at the CDW ordering vector ($L$-point) around $T_{\rm CDW}=78$~K. \ys{The intensity of} soft phonons exhibit a remarkable in-plane anisotropy, extending over a much larger momentum range along $L$-$A$ relative to $L$-$H$, which leads to diffuse scattering common among $A$V$_3$Sb$_5$. Using first-principles calculations, we find that the momentum-dependent electron-phonon coupling (EPC) is peaked at $L$ and exhibits the same in-plane anisotropy as the phonon softening. Conversely, the electronic susceptibility is not peaked at $L$ and shows the opposite in-plane anisotropy. Our findings favor momentum-dependent EPC as the driving mechanism of the CDW in KV$_3$Sb$_5$, with a CDW formation process similar to that of transition metal dichalcogenides.
\end{abstract}

\maketitle

The kagome lattice consists of corner sharing triangles, and hosts Dirac crossings, van Hove singularities, and flat bands in its electronic structure \cite{Mielke1991}. Rich emergent physics can be realized in kagome metals near the van Hove filling, whereby electronic orders such as ferromagnetism, antiferromagnetism, charge bond orders, $d$-wave Pomeranchuk instabilities, charge-density waves (CDW), and unconventional superconducting phases, have been theoretically predicted \cite{Yu2012,Kiesel2012,Wang2013,Kiesel2013}. %
Such electronic orders may be realized in the kagome metals $A$V$_3$Sb$_5$ ($A=~$K, Rb, Cs), which exhibit both CDWs and superconductivity \cite{Ortiz2019,Ortiz2020,Wilson2024}.

\begin{figure}
\includegraphics[angle=0,width=0.49\textwidth]{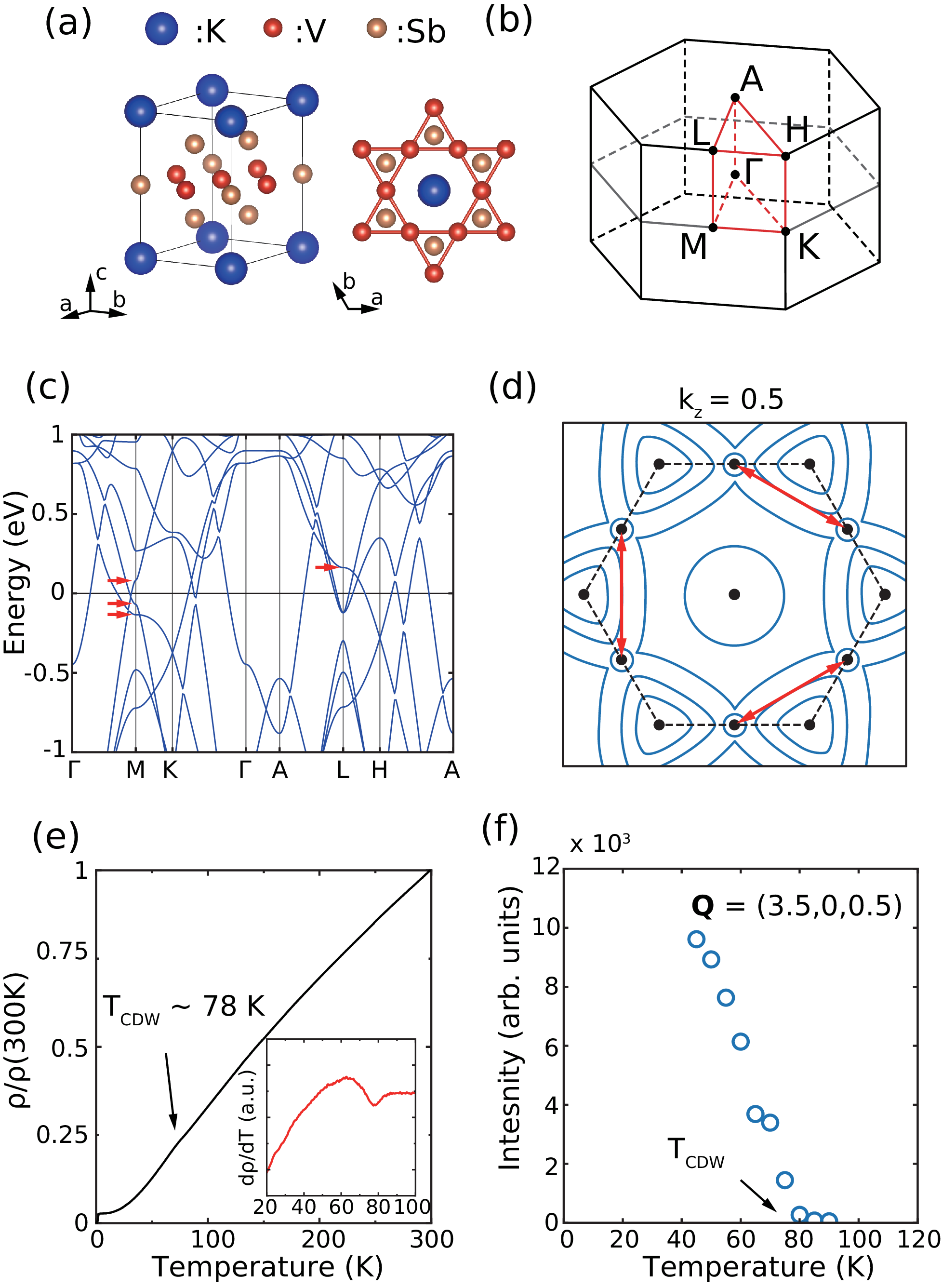}
\vspace{-12pt} \caption{\label{Figure1} (a) The crystal structure of KV$_3$Sb$_5$. (b) The Brillouin zone of KV$_3$Sb$_5$. (c) Calculated electronic structure of KV$_3$Sb$_5$, the arrows mark van Hove points near the Fermi level. (d) The Fermi surface of KV$_3$Sb$_5$ for $k_z=\frac{1}{2}$. The arrows represent the nesting between van Hove points at $M$ ($L$). (e) Normalized resistivity $\rho(T)/\rho(\rm{300~K})$ of KV$_3$Sb$_5$ single crystals. The inset shows $d\rho/d T$, with a clear anomaly at $T_{\rm CDW}=78$~K. (f) Temperature dependence of elastic x-ray scattering in {\KVS} at ${\bf Q}=(3.5,0,0.5)$.}
\vspace{-12pt}
\end{figure}

The $A$V$_3$Sb$_5$ compounds are made up of kagome lattices of V atoms coordinated by Sb octahedra, forming quasi-2D V$_3$Sb$_5$ layers separated by alkaline metal $A$ atoms [Fig.~\ref{Figure1}(a)] \cite{Wilson2024}.  CDWs with a $2\times2$ in-plane modulation appear below $T_{\rm CDW}=$~78~K, 104~K and 94~K for the K, Rb, and Cs variants, respectively. The CDW state exhibits a range of unconventional properties, including an anomalous-Hall-effect-like behavior in the absence of local moments \cite{Yang2020,Yu2021,Kenney2021}, broken time-reversal symmetry \cite{Jiang2021,Feng2021,Mielke2022}, electronic nematicity \cite{Nie2022,Grandi2023}, and switchable chiral transport \cite{Guo2022}. Superconductivity with $T_{\rm c}\approx0.9$~K for (K,Rb)V$_3$Sb$_5$ and 2.5~K for CsV$_3 $Sb$_5$ emerge inside the CDW phases at ambient pressure \cite{Ortiz2020,Yin2021,Ortiz2021a}, and persist beyond the suppression of CDW order under pressure \cite{Yu2021a,Chen2021,Du2021,Zhou2024}. Because the {\AVS} compounds share similar crystal structures, electronic structures, and CDW modulations, their CDWs likely have a common origin. Angle-resolved photoemission spectroscopy measurements show van Hove saddle points just below the Fermi level \cite{Kang2022a,Hu2022,Zhu2023},  providing evidence that the CDW is driven by the nesting between these saddle points [Figs.~\ref{Figure1}(c) and (d)] \cite{Tan2021,Park2021,Lin2021,Denner2021}. Alternatively, electron-phonon coupling (EPC) may also play an important role in driving the CDW in $A$V$_3$Sb$_5$ \cite{Luo2022,Xie2022,Liu2022,He2024,GutierrezAmigo2024,You2025}.

How a CDW forms upon cooling towards $T_{\rm CDW}$ offers important clues for its underlying mechanism. Whereas both nesting- and EPC-driven CDWs exhibit phonons that soften to zero energy at $T_{\rm CDW}$, phonon softening is typically limited to a small momentum region in the former, and is extended in the latter \cite{Chan1973,Zhu2015,Zhu2017}. On the other hand, CDWs that result from strong electron correlations, such as in the cuprates, may exhibit phonons that do not soften to zero energy \cite{LeTacon2013,He2018}.  In the case of (Cs,Rb)V$_3$Sb$_5$, phonon softening was not observed \cite{Li2021a,Subires2023}, although such a behavior is unlikely to result from electronic correlations, as $A$V$_3$Sb$_5$ are weakly correlated metals \cite{Zhou2023}. Instead, such an unconventional CDW formation has been attributed to a persistent phason gap \cite{Miao2021}, an unusually large soft phonon linewidth that obscures soft phonons \cite{GutierrezAmigo2024}, a Jahn-Teller-like instability \cite{wang2022}, a condensation of preformed CDW \cite{Park2023,Liu2025}, or strong quantum fluctuations between distinct metastable CDW states \cite{Chen2025}. 
Although the CDW transitions in (Cs,Rb)V$_3$Sb$_5$ exhibit little or no thermal hysteresis \cite{Li2021a}, they are unambiguously first-order \cite{Song2022,Zhang2024}, which may obscure the intrinsic CDW formation process in $A$V$_3$Sb$_5$. As KV$_3$Sb$_5$ exhibits a second-order CDW transition \cite{scagnoli2024}, it is particularly well-suited for probing the CDW mechanism in  $A$V$_3$Sb$_5$ via measurements of its CDW formation.

In this work, we use inelastic x-ray scattering (IXS) measurements to show that the CDW in KV$_3$Sb$_5$ forms via phonons that soften to zero energy around $T_{\rm CDW}=78$~K. The phonon softening occurs over a large region in momentum space, akin to the EPC-driven CDW in transition metal dichalcogenides \cite{Weber2011}. \ys{First-principles calculations show that the in-plane anisotropy of phonon softening matches that of the momentum-dependent EPC}, while differing significantly from the calculated electronic susceptibility. These results show that the CDW transition in KV$_3$Sb$_5$ is \ys{driven by matrix elements of the EPC}, and given the overall similarity of the $A$V$_3$Sb$_5$ compounds, such a mechanism is likely also applicable to (Cs,Rb)V$_3$Sb$_5$.

Single crystals of KV$_3$Sb$_5$ were synthesized using a self-flux method \cite{Yu2021}. Resistivity measurements show a CDW transition at $T_{\rm CDW}=78$~K and a residual resistivity ratio (RRR) of $\sim39$ [Fig.~\ref{Figure1}(e)], consistent with previous reports \cite{Ortiz2019,Ortiz2021a}. IXS measurements were carried out using the HERIX spectrometer at the Advanced Photon Source, Argonne National Laboratory, with 23.72~keV incident X-rays \cite{Said2020,Toellner2011}. The energy resolution is determined to be $\sim1.73$~meV by measuring a plexiglass standard. Momentum transfers are indexed in reciprocal lattice units (r.l.u.) of the high temperature $P6/mmm$ structure, with CDW peaks at ${\bf q}=(0.5,0.5,0.5)$ [$L$-point in Fig.~\ref{Figure1}(b)] and equivalent positions. First-principles calculations were performed using the \texttt{Quantum ESPRESSO} package \cite{Giannozzi_2017}, using the Perdew, Burke, and Ernzerhof parameterization of the generalized gradient approximation \cite{GGA}. The lattice parameters were optimized with a $\Gamma$-centered $12\times12\times8$ $\mathbf{k}$-mesh, and different Gaussian smearings were used to simulate different electronic temperatures \cite{Souliou2022,Korshunov2023,Cao2023}. The phonon spectrum was obtained using density-functional perturbation theory (DFPT) with a $4\times4\times2$ $\mathbf{q}$-mesh \cite{DFPT}. The electron-phonon coupling was calculated using the \texttt{EPW} \cite{EPW} package, and the bare electron susceptibility was calculated using a symmetrized Wannier Hamiltonian \cite{wannier90,WannSymm}. See the Supplemental Material for further experimental and computational details \cite{SI}.

\begin{figure}
\includegraphics[angle=0,width=0.49\textwidth]{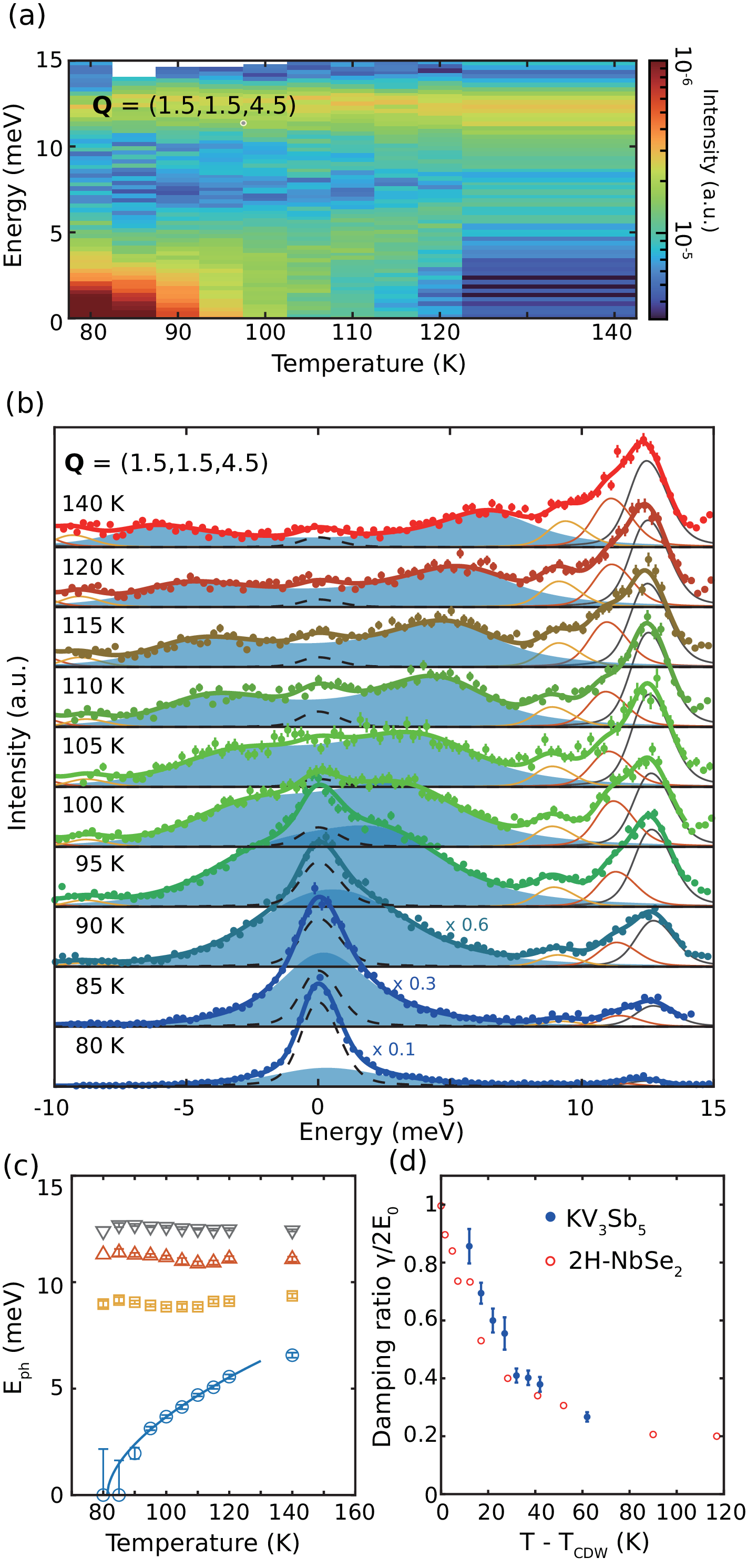}
    \vspace{-12pt} 
    \caption{
    (a) Color map of scattering intensities for ${\bf Q}=(1.5,1.5,4.5)$ at various temperatures and energies. (b) IXS spectra at various temperatures (vertically shifted), with the corresponding fits shown as solid lines. The resolution-limited elastic peak and phonon modes are shown as dashed and solid lines, respectively. The soft phonon mode is modeled using a DHO, and is represented by the shaded area. The scans at 80~K, 85~K and 90~K are scaled by the factors indicated in the figure. In fitting the 80~K and 85~K data, some parameters were constrained \cite{SI}. (c) Temperature dependence of the fitted phonon energies $E_{\rm ph}$. The solid blue line is a power law fit. (d) Damping ratio $\gamma/(2E_0)$ of the soft mode in KV$_3$Sb$_5$, compared with the results for $2H$-NbSe$_2$ \cite{Weber2011}.
    }
    \vspace{-12pt}
    \label{Figure2}
\end{figure}

Elastic scattering measured at ${\bf Q}=(3.5,0,0.5)$ reveals a gradual onset of the CDW peak intensity below $T_{\rm CDW}$, demonstrating the second-order character of the CDW transition in KV$_3$Sb$_5$ [Fig.~\ref{Figure1}(f)], consistent with previous results \cite{scagnoli2024}. IXS scans at various temperatures for ${\bf Q}=(1.5,1.5,4.5)$ (an $L$-point) are shown in Fig.~\ref{Figure2}(a), revealing a band of scattering with a maximum intensity at $\sim13$~meV that does not change significantly with temperature, and a mode that softens from $\sim6.5$~meV towards the elastic line, upon cooling from 140~K to $T_{\rm CDW}$. The electron-phonon coupling $\lambda$ is related to the soft mode energy $E_{\rm sm}$ and the amplitude mode energy $E_{\rm am}$ via $\lambda=(E_{\rm am}/E_{\rm sm})^2$. For KV$_3$Sb$_5$, our measurements show $E_{\rm sm}\approx 6.5$~meV at 140~K, and Ref.~\cite{Wu2022} shows $E_{\rm am}\approx12$~meV, which lead to $\lambda\approx3.4$. This value is reasonably large, and similar to the value inferred for CsV$_3$Sb$_5$ \cite{Liu2022}.

To quantitatively assess the temperature-evolution of phonons in Fig.~\ref{Figure2}(a), the IXS spectrum at each temperature is presented in Fig.~\ref{Figure2}(b), which are fitted to the sum of a resolution-limited elastic peak, three resolution-limited phonon modes with energies between $\sim9$~meV and $\sim13$~meV, and a damped harmonic oscillator (DHO) that characterizes the low-lying mode, after convolution with the instrumental resolution. The DHO takes the form:
\begin{equation*}
S(E)=\frac{1}{1-\exp(-\frac{E}{k_{\rm B}T})}\frac{A}{E_0}\frac{2\gamma E E_0}{\pi[(E^2-E_0^2)^2+(E\gamma)^2]},
\label{DHO}
\end{equation*}
where $A$ is the intensity, $E_0$ is the undamped phonon energy, and $\gamma$ is the damping rate (peak width for $\gamma\ll E_0$) \cite{Lamsal2016,Song2023,Cao2023}. The damping rate $\gamma$ renormalizes the phonon energy from $E_0$ to $E_{\rm ph}=\sqrt{E_0^2-\gamma^2/4}$, with $E_{\rm ph}\rightarrow0$ as $\gamma/(2E_0)\rightarrow1$. 
Whereas $E_{\rm ph}$ of the three modes between $\sim9$~meV and $\sim13$~meV are essentially temperature-independent, the low-lying mode exhibits a strong temperature dependence, softening towards zero energy upon cooling to $T_{\rm CDW}$.
Fitting a power law behavior $E_{\rm ph}=[(T-T_0)/T_0]^\delta$ yields \ys{$T_0=81(3)$~K and $\delta=0.56(8)$}, in good agreement with a mean-field softening of phonons towards zero energy at $T_{\rm CDW}$, a behavior also found in $2H$-NbSe$_2$ \cite{Weber2011} . Also closely resembling $2H$-NbSe$_2$ \cite{Weber2011}, we find $\gamma/(2E_0)$ of KV$_3$Sb$_5$ approaches unity upon cooling towards $T_{\rm CDW}$ [Fig.~\ref{Figure2}(d)]. 

%

\begin{figure}
    \includegraphics[angle=0,width=0.49\textwidth]{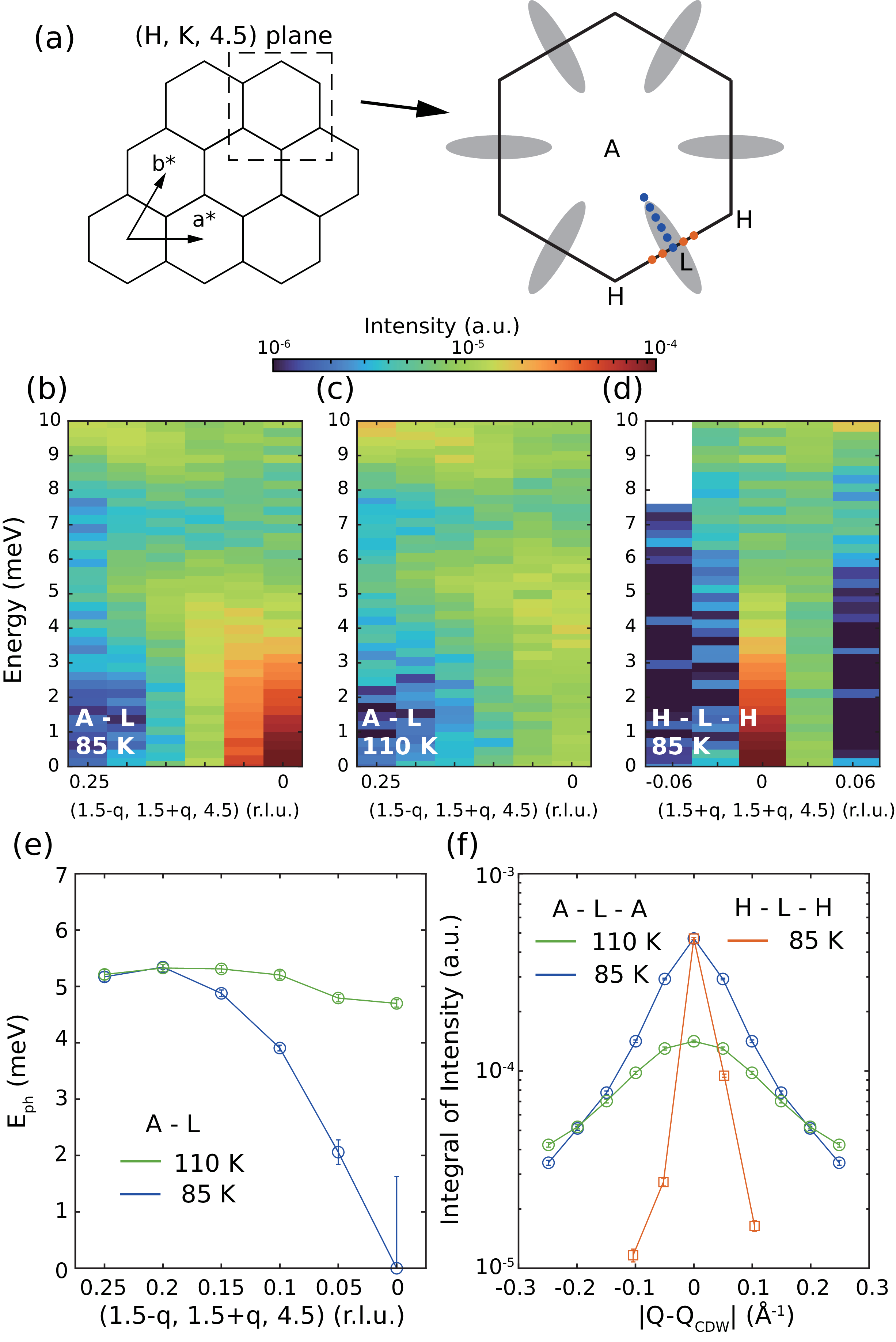}
    \vspace{-12pt} 
    \caption{
    (a) Schematic of the reciprocal space probed in this work, with the dashed area zoomed in on the right. Dots indicate reciprocal space positions where IXS scans were performed. Color maps of IXS scans along $A-L$ at (b) 85~K and (c) 110~K. (d) Color map of IXS scans along $H-L-H$ at 85~K. See Ref.~\cite{SI} for detailed IXS scans and fits corresponding to (b)-(d). (e) Dispersion of phonon energies $E_{\rm ph}$ along $A-L$ at the two temperatures. (f) Integral of inelastic scattering intensities for points along $A-L-A$ and $H-L-H$, obtained from data in (b)-(d) after a resolution-limited elastic peak is subtracted. The data for $L-A$ is mirrored from that of $A-L$. 
    }
    \vspace{-12pt}
    \label{Figure3}
\end{figure}

To explore the \ys{momentum dependence of soft phonons}, points along $A-L$ [${\bf Q}$ from (1.25, 1.75, 4.5) to (1.5, 1.5, 4.5)] and $H-L-H$ [${\bf Q}$ from (1.44, 1.44, 1.5) to (1.56, 1.56, 4.5)] are selected for IXS scans [dots in Fig.~\ref{Figure3}(a)]. \ys{Soft phonons are} observed over a large momentum range along $A-L$, with significantly softer phonons at 85~K relative to 110~K [Figs.~\ref{Figure3}(b) and (c)]. The $A-L$ phonons at both temperatures are well described by the model in Fig.~\ref{Figure2}(b), with the momentum dependence of $E_{\rm ph}$ for the low-lying mode shown in Fig.~\ref{Figure3}(e), revealing phonon softening over an extended momentum range of $\sim0.25$~\AA$^{-1}$ along $A-L$ ($\sim0.5$~\AA$^{-1}$ along $A-L-A$). \ys{On the other hand, soft phonons are} limited to a small momentum range along $H-L-H$, and are already undetectable 0.1~\AA$^{-1}$ away from $L$ (\ys{within 0.2~\AA$^{-1}$ along $H-L-H$}). This in-plane anisotropy \ys{of the soft phonon intensity} has two contributions: (1) an anisotropy in the phonon dispersion, and (2) an anisotropy in the phonon structure factor \cite{SI}.

To directly visualize the in-plane anisotropy of \ys{the soft phonon intensity}, the inelastic scattering in Figs.~\ref{Figure3}(b)-(d) are integrated from -7~meV to 7~meV and the resulting $I_{\rm int}$ are compared in Fig.~\ref{Figure3}(f), after subtracting resolution-limited elastic peaks from the data. The results from $A-L$ are mirrored to $L-A$. Since the soft phonon mode dominates inelastic scattering in this window, $I_{\rm int}$ is a proxy for the intensity of soft phonons. Compared to intensity at the $L$-point (${\bf Q}_{\rm CDW}$), $I_{\rm int}$ drops by $\sim30$\% for $|{\bf Q}-{\bf Q}_{\rm CDW}|=0.1~${\AA} along $L-H$, whereas it drops by $\sim3$\% for \ys{a similar displacement from ${\bf Q}_{\rm CDW}$ along $L-A$}. The highly anisotropic \ys{intensity of soft phonons} will give rise to thermal diffuse scattering with a similar anisotropy, as schematically represented by the shaded ellipses in Fig.~\ref{Figure3}(a). Diffuse scattering with such an in-plane anisotropy is found in CsV$_3$Sb$_5$ and RbV$_3$Sb$_5$ \cite{Subires2023}, which demonstrates that the {\AVS} compounds exhibit a common anisotropic diffuse scattering above $T_{\rm CDW}$, pointing to a common underlying CDW mechanism.

\begin{figure}
    \includegraphics[angle=0,width=0.49\textwidth]{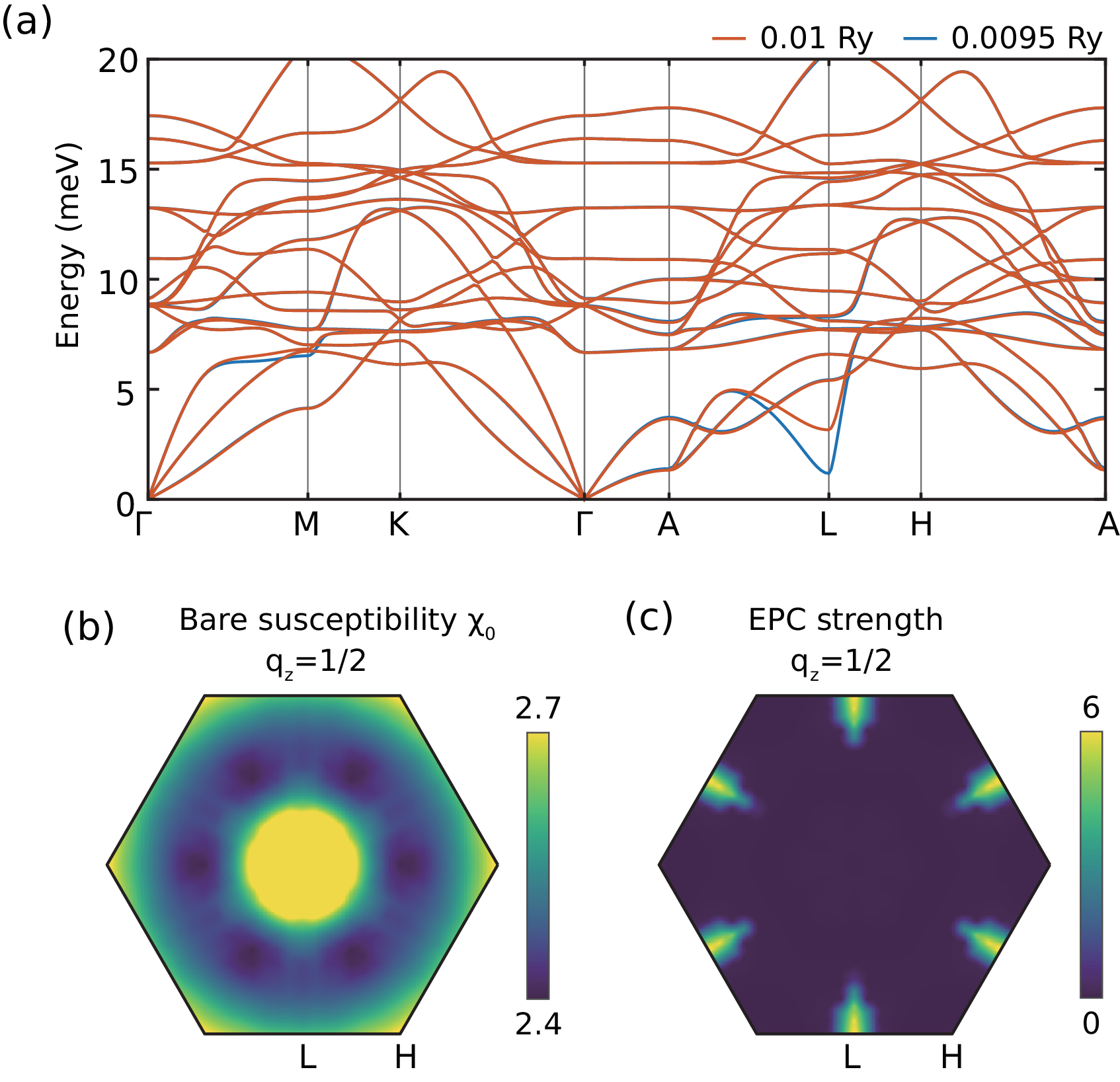}
    \vspace{-12pt} 
    \caption{
    (a) Calculated phonon dispersions in {\KVS}, \ys{for Gaussian smearings of 0.01~Ry and 0.0095~Ry. Similar results for 0.1~Ry and 0.001~Ry are shown in the Supplemental Materials \cite{SI}.}  (b) The bare susceptibility $\chi_0(q)$ in the $q_z=\frac{1}{2}$ plane. The strong peak at $A$ is due to self-nesting in a quasi-2D system. (c) ${\bf q}$-dependent EPC $\lambda_{\bf q}$ for the soft phonon mode in the $q_z=\frac{1}{2}$ plane, computed at \ys{0.01~Ry}.}
    \vspace{-12pt}
    \label{Figure4}
\end{figure}

To shed light on the CDW mechanism in {\KVS}, we compute the electronic structure and phonon dispersion of {\KVS} using first-principles calculations. The electronic structure of {\KVS} is shown in Fig.~\ref{Figure1}(c), with several van Hove points close to the Fermi level (indicated by arrows), consistent with previous results \cite{kato2022,Luo2022}. The calculated phonon dispersions are shown in Figs.~\ref{Figure4}(a), using different Gaussian smearings to simulate \ys{temperatures approaching $T_{\rm CDW}$ from above} \cite{Souliou2022,Cao2023,Korshunov2023}. 
\ys{The calculated longitudinal acoustic (LA) mode exhibits a clear dip at $L$ and an in-plane anisotropy around $L$: at 0.01~Ry, the LA phonon dispersion along $L-H$ is much stiffer compared to $L-A$, and upon decreasing temperature to 0.0095~Ry, the softening of the LA phonon is more prominent along $L-H$ compared to $L-A$.} We note that a similar anisotropy in phonon dispersion around $L$ is also obtained in calculations that simulate finite temperatures by considering anharmonic effects \cite{Ptok2022}.
\ys{For comparison, the LA phonon at $L$ becomes imaginary at a low temperature (0.001~Ry), and the dip at $L$ disappears upon increasing the Gaussian smearing to 0.1~Ry \cite{SI}.}



\ys{In order to determine the mechanism that accounts for the in-plane anisotropy of phonon softening in the calculated phonon dispersion, the bare electronic susceptibility (with form factor) $\chi_0({\bf q})$ and the ${\bf q}$-dependent EPC $\lambda_{\bf q}$ for the LA phonon are computed at $0.01$~Ry} and shown in Figs.~\ref{Figure4}(b) and (c). 
$\chi_0({\bf q})$ exhibits hexagonal ridges around the zone boundary, these ridges are more intense at $H$ than at $L$, and are more extended along $L-H$ than $L-A$. The fact that $\chi_0({\bf q})$ is not maximized at $L$ and has an in-plane anisotropy opposite to \ys{the calculated anisotropy of phonon softening in Fig.~\ref{Figure4}(a)}, suggests that the CDW in {\KVS} is unlikely to originate from Fermi surface nesting. This is consistent with previous calculations of the Lindhard response function in {\AVS} \cite{Kaboudvand2022}. On the other hand, the EPC $\lambda_{\bf q}$ peaks at $L$, with the peak being more elongated along $A-L$ relative to $H-L$, consistent with ys{the calculated phonon softening in Fig.~\ref{Figure4}(a)}.  These findings strongly favor ${\bf q}$-dependent EPC \ys{matrix element effects} over nesting as the dominant driver of the CDW in {\KVS}, and are corroborated by low temperature \ys{(0.001~Ry)} calculations that use pressure to suppress the CDW \cite{SI}. 


Our experimental and theoretical results show that the CDW in {\KVS} forms via phonon softening towards zero energy, with the soft phonons exhibiting a prominent in-plane anisotropy that is consistent with the momentum-dependent EPC. The CDW formation in {\KVS} exhibits phonon softening over an extended momentum region [Fig.~\ref{Figure3}(e)], a mean-field temperature dependence of $E_{\rm ph}$ [Fig.~\ref{Figure2}], and a phonon damping rate $\gamma/(2E_0)$ that gradually increases towards unity at $T_{\rm CDW}$ [Fig.~\ref{Figure2}(d)], similar to the EPC-driven CDW of $2H$-NbSe$_2$ \cite{Weber2011}, further supporting an EPC-driven CDW in {\KVS}. The common diffuse scattering [Fig.~\ref{Figure3}(a)] then suggests EPC to be the common mechanism of the CDWs in {\AVS}, although electronic effects \cite{Eiter2012,Enzner2025} may be also at play and contribute to the unconventional properties reported for these materials.

Our findings raise the question as to why phonon softening is not observed in (Cs,Rb)V$_3$Sb$_5$ \cite{Li2021a,Subires2023}, for which there are several potential contributing factors. As the CDW transitions in (Cs,Rb)V$_3$Sb$_5$ are first-order \cite{Song2022,Zhang2024}, the transition may occur before the corresponding phonons significantly soften, which contrasts with {\KVS} that shows a second-order CDW transition \cite{scagnoli2024}. Compared to {\KVS} that shows a single CDW phase in its temperature-pressure phase diagram \cite{Du2021,Kautzsch2023}, CsV$_3$Sb$_5$ has CDWs modulated by $(\frac{1}{2},\frac{1}{2},\frac{1}{2})$ and $(\frac{1}{2},\frac{1}{2},\frac{1}{4})$ at ambient pressure \cite{Hu2022a,Stahl2022,Xiao2023}, and a pressure-induced $(0,\frac{3}{8},\frac{1}{2})$ CDW under pressure \cite{Zheng2022,Stier2024}. In RbV$_3$Sb$_5$, there may also be a pressure-induced CDW distinct from the ambient pressure CDW, although its transport signatures are much weaker compared to CsV$_3$Sb$_5$ \cite{Wang2021,Du2022}.  Such a coexistence of distinct CDW instabilities close in energy may frustrate the system and hinder the softening of phonons at a particular wave vector \cite{Chen2025}. Nonetheless, as the CDW-related diffuse scattering in (Cs,Rb)V$_3$Sb$_5$ \cite{Subires2023} are similar to that in {\KVS}, the presence of soft phonons in these compounds cannot be excluded. As the soft phonons may be weak in some Brillouin zones [e.g. soft phonons in {\KVS} were not detected at ${\bf Q}=(3.5,0,0.5)$], a comprehensive search of soft phonons across multiple zones in (Cs,Rb)V$_3$Sb$_5$ will be helpful to elucidate whether phonon softening is common to the CDW formation of {\AVS}.

Given the role of EPC in driving the CDW of {\AVS}, it may also play a role in the superconductivity that emerges from the CDW state \cite{Jiang2023,You2025}. In such a scenario, the pairing symmetry is expected to be nodeless, for which there has been considerable evidence \cite{Duan2021,Zhou2024,Mine2024,Kaczmarek2025}, although there are also reports of unconventional pairing \cite{Guguchia2023,Deng2024,Zhao2024}. The role of the lattice (and thus EPC) in the superconductivity of {\AVS} is also underscored by the tuning of $T_{\rm c}$ via structural crossovers or transitions under hydrostatic pressure \cite{Tsirlin2022,Yu2022,Du2022a}. Furthermore, our calculations show that EPC reasonably accounts for superconductivity of {\KVS}, when its CDW is destabilized under pressure \cite{SI}.


In conclusion, we find that the CDW in {\KVS} forms via phonon softening towards zero energy around $T_{\rm CDW}$, with soft phonons \ys{observed} over an extended momentum range along $A-L$ but a narrow range along $H-L$. \ys{The phonon softening of {\KVS} is} reproduced in first-principles calculations, \ys{and is shown to arise} from a momentum-dependent EPC. These results evidence a conventional EPC mechanism of the CDW formation in {\KVS}, similar to that in transition metal dichalcogenides, which likely also applies to other members of the {\AVS} family.

The work at Zhejiang University was supported by the National Key R\&D Program of China (No. 2022YFA1402200, 2023YFA1406303), the National Natural Science Foundation of China (No. 12350710785, 12274363, 12274364, 12304175, 12494592), and the Fundamental Research Funds for the Central Universities (Grant No. 226-2024-00068). This research used resources of the Advanced Photon Source, a U.S. Department of Energy (DOE) Office of Science user facility operated for the DOE Office of Science by Argonne National Laboratory under Contract No. DE-AC02-06CH11357.

Note: After the submission of this work, soft phonons were also reported for CsV$_3$Sb$_5$ \cite{McGuinness2025}, pointing to a common soft-phonon CDW formation in the {\AVS} compounds.


\bibliography{K135.bib}

\end{document}